\documentstyle[12pt,lathuile]{article}
\begin{document}

\newcommand{\perc}      {~\mbox{\hspace*{-0.3em}\%}}
\newcommand{\Gev}       {\mbox{${\rm GeV}$}}
\newcommand{\Gevsq}     {\mbox{${\rm GeV}^2$}}
\newcommand{\Fem}       {\mbox{$F_2^{em}$}}
\newcommand{\Fint}      {\mbox{$F_2^{int}$}}
\newcommand{\Fwk}       {\mbox{$F_2^{wk}$}}
\newcommand{\Ft}        {\mbox{$F_2$}}
\newcommand{\Fd}        {\mbox{$F_3$}}
\newcommand{\FL}        {\mbox{$F_{_{L}}$}}
\newcommand{\qsd}       {\mbox{${Q^2}$}}
\newcommand{\x}         {\mbox{{\it x}}}
\newcommand{\y}         {\mbox{{\it y}}}
\newcommand{\ye}        {\mbox{${y_{e}}$}}
\newcommand{\smallqsd}  {\mbox{${q^2}$}}
\newcommand{\ra}        {\mbox{$ \rightarrow $}}
\newcommand{\ygen}      {\mbox{${y_{gen}}$}}
\newcommand{\yjb}       {\mbox{${y_{_{JB}}}$}}
\newcommand{\yda}       {\mbox{${y_{_{DA}}}$}}
\newcommand{\qda}       {\mbox{${Q^2_{_{DA}}}$}}
\newcommand{\qjb}       {\mbox{${Q^2_{_{JB}}}$}}
\newcommand{\ypt}       {\mbox{${y_{_{PT}}}$}}
\newcommand{\qpt}       {\mbox{${Q^2_{_{PT}}}$}}
\newcommand{\ypr}       {\mbox{${y_{(1)}}$}}
\newcommand{\yprpr}     {\mbox{${y_{(2)}}$}}
\newcommand{\ypps}      {\mbox{${y_{(2)}^2}$}}
\newcommand{\gammah}    {\mbox{$\gamma_{_{H}}$}}
\newcommand{\gammahc}   {\mbox{$\gamma_{_{PT}}$}}
\newcommand{\gap}       {\hspace{0.5cm}}
\newcommand{\gsim}      {\mbox{\raisebox{-0.4ex}{$\;\stackrel{>}{\scriptstyle \sim}\;$}}}
\newcommand{\lsim}      {\mbox{\raisebox{-0.4ex}{$\;\stackrel{<}{\scriptstyle \sim}\;$}}}
\newcommand{\ptrat}     {\mbox{$\frac{p_{Th}}{p_{Te}}$}}
\newcommand{\yjbrat}    {\mbox{$\frac{y_{_{JB}}}{y_{gen}}$}}
\newcommand{\ydarat}    {\mbox{$\frac{y_{_{DA}}}{y_{gen}}$}}
\newcommand{\yptrat}    {\mbox{$\frac{y_{_{PT}}}{y_{gen}}$}}
\newcommand{\yprrat}    {\mbox{$\frac{y_{( 1)}}{y_{gen}}$}}
\newcommand{\yprprrat}  {\mbox{$\frac{y_{(2)}}{y_{gen}}$}}
\newcommand{\yerat}     {\mbox{$\frac{y_{e}}{y_{gen}}$}}
\newcommand{\yptye}      {\mbox{$\frac{y_{_{PT}}}{y_{e}}$}}
\newcommand{\qptrat}    {\mbox{$\frac{Q^2_{PT}}{Q^2_{gen}}$}}
\newcommand{\qdarat}    {\mbox{$\frac{Q^2_{DA}}{Q^2_{gen}}$}}
\newcommand{\qerat}     {\mbox{$\frac{Q^2_{e}}{Q^2_{gen}}$}}
\newcommand{\qprprrat}  {\mbox{$\frac{Q^2_{(2)}}{Q^2_{gen}}$}}
\title{ 
STRUCTURE FUNCTIONS AND LARGE {\boldmath $Q^2$} CROSS SECTIONS AT HERA
}
\author{
Giuseppe Iacobucci \thanks{\ \ Talk given on behalf of the H1 and ZEUS collaborations at the Workshop on Results and 
Perspectives in Particle Physics, La Thuile, Italy, February 2000}        \\
{\em INFN-Sezione di Bologna, Via Irnerio 46, I-40126 Bologna, Italy}
}
\maketitle
\baselineskip=14.5pt
\begin{abstract}
The data collected with the H1 and ZEUS detectors
during the running period 1994-99 are used to give 
an experimental review on the proton structure functions 
and the 
neutral current and charged current large $Q^2$ cross sections
in $e^\pm p$
scattering at HERA.
\end{abstract}
\baselineskip=17pt
\newpage

\pagestyle{plain}
\footskip 1.2cm

\section{Foreword}
With the advent of the HERA $e^{\pm} p$ collider at DESY, 
enormous progress in the measurement of the proton structure has been made.
The large HERA $ep$ centre-of-mass energy $\sqrt{s} = 300 - 318$ GeV,
obtained with a lepton-beam energy $E_e = 27.5$ GeV and proton-beam energies 
$E_p = 820$ GeV until 1997 and $E_p = 920$ GeV since 1998,
allowed the two collider experiments H1 and ZEUS 
to measure \Ft~up to virtualities of the exchanged boson $Q^2 \sim 10^5$ GeV$^2$ and down to Bjorken $x \sim 10^{-6}$.
These measurements 
constitute an extention by more than two orders of magnitude 
of the $(x, Q^2)$ range in which we have knowledge of the proton structure
(see Fig. 1).
\begin{figure}[h]
 \vspace{11.0cm}
\includegraphics{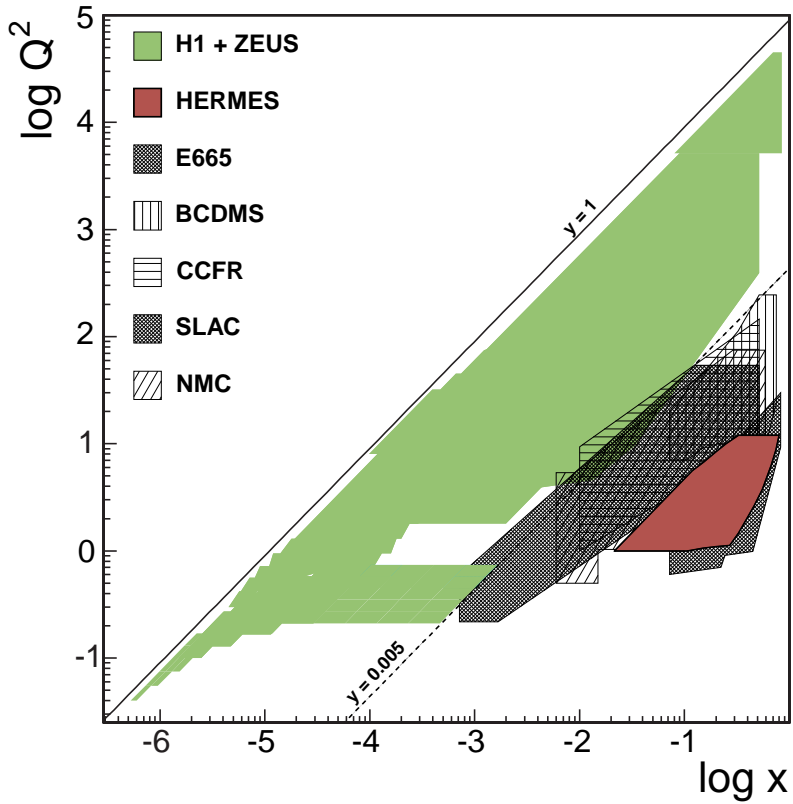}
 \caption{\it
The $(x,Q^2)$ kinematic plane, showing the range covered by 
the HERA and the fixed target measurements.
}
\end{figure}
The study of $e^\pm p$ interactions is important throughout the above
kinematic range, not only because of the structure functions,
which allow a calculation of the expected
rates at the LHC and an estimation of the $\nu N$ cross sections 
for ultra-high-energy neutrinos\cite{uhn} 
($E_\nu \sim 10^{12}$ GeV) from active galactic nuclei 
and $\gamma$-ray bursts, 
but also because of several theoretical issues 
still open in this field:
\begin{itemize}
\setlength{\itemsep}{-2pt}
\item[--] it is of primary importance to study the transition from the
photoproduction ($Q^2 \sim 0$) to the deep inelastic scattering 
(DIS, $Q^2 \gsim$ few GeV$^2$) regime
to see at which value of \qsd~perturbative QCD (pQCD) 
begins to dominate\cite{mandy};
\item[--] the strong rise of $F_2$ measured in DIS at HERA 
for $x \rightarrow 0$
is a well established and important fact. 
However, it is imperative to measure
\Ft~with higher and higher precision over a wide kinematic range,
in order to address  
the issues of parton saturation at very small $x$\cite{sat}
and QCD evolution 
(the BFKL evolution\cite{bfkl},
which ought to be important when terms in $\ln \frac{1}{x}$
become large, has yet to be observed experimentally);
\item[--] 
as will be shown in the following,
the study of electroweak (EW) physics at HERA\cite{ew} is just beginning.
Large luminosities both with electrons and positrons, as well
as polarised beams, are needed to perform these studies, 
which have always been 
one of the main aims of HERA;
\item[--] the exploration of new kinematic regimes
may reveal new physics.
The breakdown of the Standard Model
may manifest itself as deviations of the measured cross sections 
from the predictions at very large $(x,Q^2)$.
\end{itemize}
In this document, an experimental review  will be given 
of the knowledge of proton structure functions and
$ep$ cross sections at large $Q^2$ using
the H1 and ZEUS results, 
obtained with the $\sim 70~pb^{-1}$ of $e^+p$ 
and $\sim 15~pb^{-1}$ of $e^-p$ collisions that each experiment
has collected until the end of 1999.

\section{\bf Kinematics}
\label{KINREC}

In inclusive deep inelastic scattering,
$e(k) + p(P) \rightarrow e (k^\prime) + X$,
the proton structure functions are expressed in terms 
of the negative of the four-momentum transfer squared:
\[
Q^2  =  - q^2 =  -(k-k^\prime)^2
\]
and of the Bjorken $x$:
\[
x    =  \frac{Q^2}{2P \cdot q}  
\]
where $k$ and $P$  are the four-momenta of the 
incoming particles and $k^\prime$ is the four-momentum of the scattered 
lepton.
The fraction of the lepton energy transferred to the proton
in its rest frame is $y = Q^2/(sx)$.
The ZEUS and H1 detectors measure the energies and angles of the
scattered lepton and hadronic system.  
These four independent quantities
over-constrain the kinematic
variables $x$ and $Q^2$ (or, equivalently, $y$ and $Q^2$).
In order to optimise the reconstruction of the kinematic variables,
the two collaborations use different methods, 
dictated by the characteristics of their 
detectors\cite{mandy}.
 
\section{The proton structure functions }
\label{section:rew}
In inclusive deep inelastic scattering the double differential cross section for 
the exchange of a neutral current (NC)
is given by:
\begin{equation}
\frac{d^2\sigma^{e^{\pm} p}_{NC}}{dx\;d\qsd} = \frac{2\pi \alpha^2}{xQ^4}
\left[Y_+ \Ft (x,\qsd )-y^2\FL (x,\qsd ) \mp Y_{-}x\Fd (x,\qsd )\right]
(1+\delta_r(x,\qsd )) \label{ncxsect}
\end{equation}
where $Y_\pm = 1 \pm (1-y)^2$, $\alpha$ is the EW coupling constant
and $\delta_r$ is the EW radiative correction.
In leading order QCD and for longitudinally unpolarized beams,
the longitudinal structure function $F_L = 0$, 
while the structure functions $F_2$ and $xF_3$ 
are expressed as sums over the quark flavor $f$ of the product of the EW
quark couplings, $A_f$ and $B_f$, and the quark momentum distributions 
in the proton $q_f (x,Q^2)$:
\[
F_2 = x \sum\limits_{f}\, A_f^2 \cdot (q_f (x,Q^2) + \bar q_f (x,Q^2)
\]
\[
xF_3 = x \sum\limits_{f}\, B_f^2 \cdot (q_f (x,Q^2) - \bar q_f (x,Q^2)
\]
where, at low $Q^2$, $A_f$ reduces to the quark electric charge.
The parity violating term $xF_3$,
due to $Z^0$ exchange, becomes relevant only for
$Q^2 \simeq M_{Z^0}^2$,
where $M_{Z^0}$ is the mass of the $Z^0$ boson. 
Beyond the leading order in QCD,
the emission of gluons
allows longitudinally polarised  photons to be absorbed by 
spin $\frac{1}{2}$ quarks.
Therefore $F_L$ becomes non-zero and can be written as a function of $F_2$
and the gluon momentum distribution $xg(x,Q^2$):
\begin{equation}
F_L=\frac{\alpha_s}{4\pi} x^2 \int \frac{dz}{z^3} 
\left[ \frac{16}{3} F_2 + 8 \sum e_i^2 \left(1-\frac{x}{z}\right) zg \right].
\label{fl}
\end{equation}
The effect of $F_L$ on the cross section 
is negligible at small values of $y$ and 
becomes substantial at large $y$.

\subsection{High precision $F_2$ measurements at very small $(x,Q^2)$}

During 1997, the beam pipe tracker\cite{bpt} (BPT),
a tracking device based on silicon microstrip technology, 
was installed in the
ZEUS detector in front of the existing beam pipe calorimeter (BPC), a small 
electromagnetic sampling calorimeter positioned at small 
scattered-lepton angles to measure the energy of the scattered lepton
in low-$Q^2$ events. 
The installation of the BPT improved the 
measurement of the scattered-lepton angle and allowed 
the fiducial range of the BPC to be increased, thus extending the
kinematic range of the $F_2$ measurement at very small $(x,Q^2)$
to $0.045 < Q^2 < 0.65$ GeV$^2$ and
$6 \cdot 10^{-7} < x < 1 \cdot 10^{-3}$.
The measured $F_2$ as a function of $x$ for several $Q^2$ values,
obtained with $3.9~pb^{-1}$ of 
$e^+ p$ collisions collected with the BPT,
are shown in Fig. 2. 
 \begin{figure}[t]
\centering
 \vspace{13cm}
\includegraphics{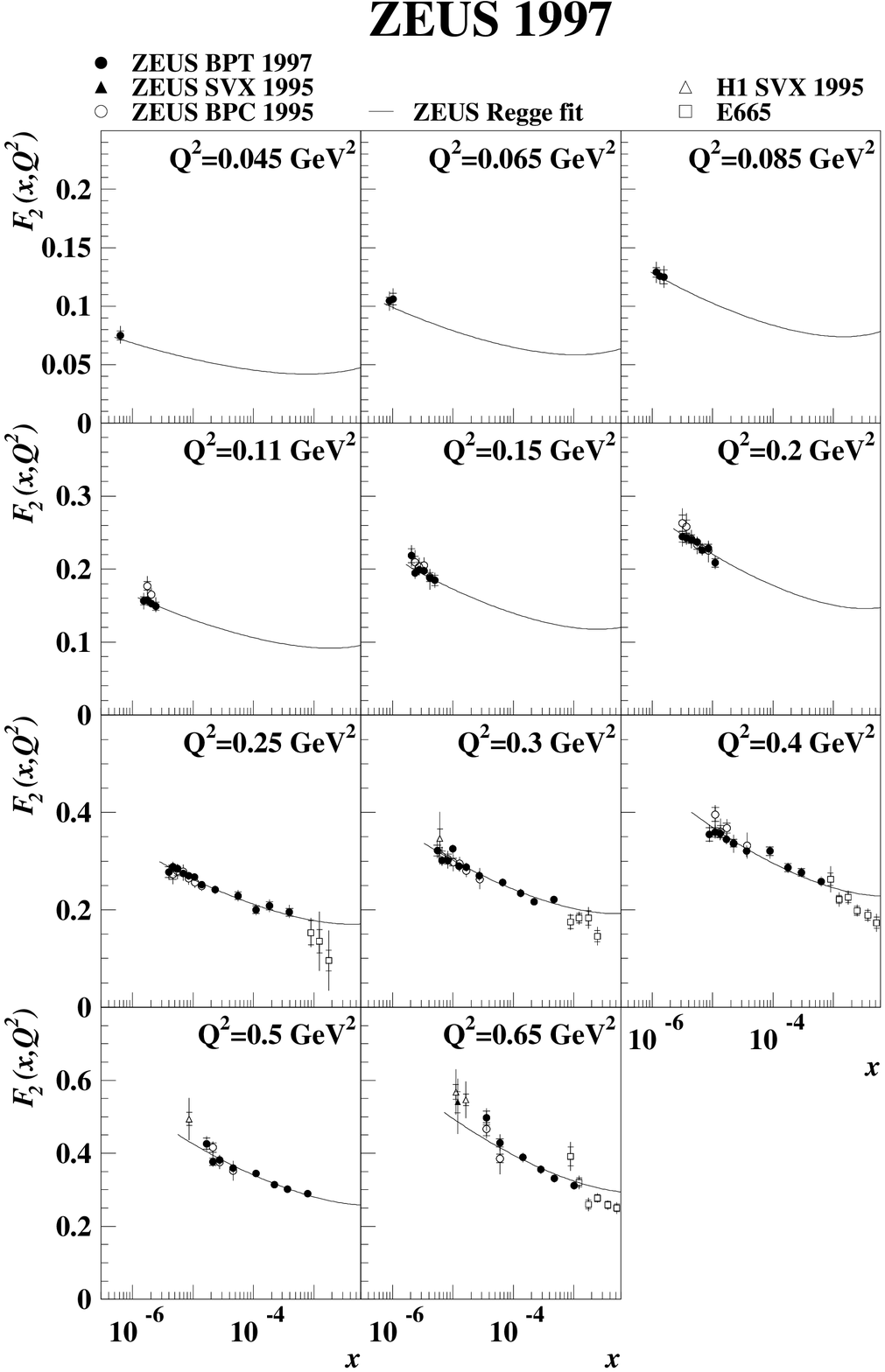}
 \caption{\it
   The   HERA measurement of $F_2$ at very small $x$ and $Q^2$.
E665 data are also shown. 
The lines are the result of a fit to the data based on 
Regge phenomenology\protect\cite{bpt}.
    \label{f2bpt} }
\end{figure}
The typical uncertainties are $\pm 2.6 \% (stat.)$ 
and $ \pm 3.3 \% (syst.)$.
The rise of $F_2$ for $x \rightarrow 0$ is measured to become slower 
as $Q^2$ decreases,
and can be described by Regge theory 
with a constant logarithmic slope.
The dependence of $F_2$ on $Q^2$ is stronger than at higher $Q^2$ values, 
approaching, at the lowest $Q^2$ of this measurement, 
a region where $F_2$ becomes nearly proportional to $Q^2$\cite{bpt}.

\subsection{$F_2$, its derivatives and the QCD NLO fit at medium $Q^2$}

\subsubsection{\it $F_2$ measurements at medium $Q^2$ at HERA: 
precision data}

The data samples collected during 1996-97 with the H1 and ZEUS detectors, 
each corresponding to approximately $40~pb^{-1}$,
made possible a precise measurement of \Ft~at medium $Q^2$.
  \begin{figure}[t]
\vspace{13.5cm}
\includegraphics{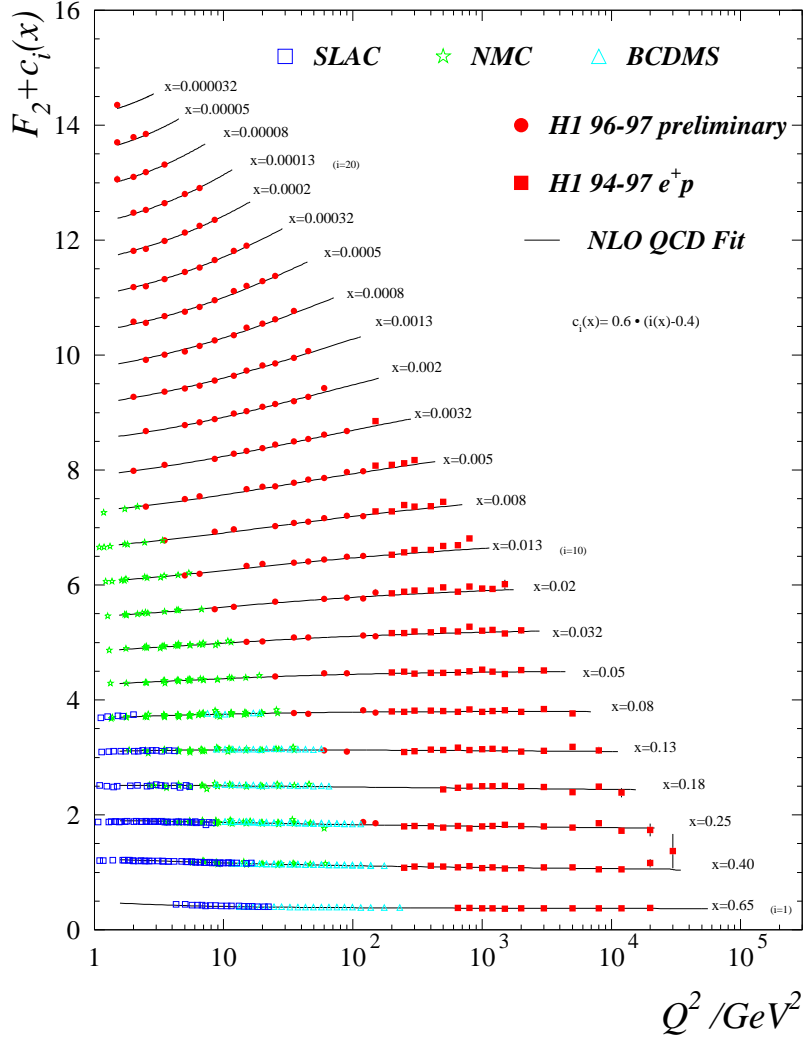}
 \caption{\it
The H1 measurement of 
$F_2$ vs. $Q^2$ in bins of $x$; fixed-target data and 
the H1 pQCD NLO fit are also shown.
In order to improve visibility,
the data and the fit are multiplied by the $c_i(x)$
factors as shown in the Figure.
    \label{f2} }
\end{figure}
This improvement was possible both because of the higher 
statistics and a better knowledge of the systematic uncertainties.
The latter was partly due to the installation of new detector 
components, such as the H1 backward silicon detector. Typical uncertainties 
of approximately $1 \%$ statistical and $2 - 3 \%$ systematic
were achieved, thus approaching the precision of the 
fixed-target experiments.
An overview of the \Ft~measurements\cite{max,niels}
 is given in Fig.~3.
The HERA and the fixed-target data agree in the region of overlap.
A pQCD next-to-leading-order (NLO) fit  (see Section 3.2.3)
using the DGLAP evolution equations\cite{dglap}
describes the data over the full range of the measurement.
Scaling violations are well described by the fit. 
There is no indication that the HERA data
require any $(\ln \frac{1}{x})^n$ BFKL-type correction terms to the 
standard DGLAP evolution.

\subsubsection{\it Derivatives of $F_2$}
As pointed out in the literature\cite{bartels}, the slopes
$d\ln F_2/d\ln \frac{1}{x}$ and
$dF_2/d\ln Q^2$
contain a lot of information.
At fixed $Q^2$ and at small $x$ the behaviour of
$F_2$ can be characterised by $F_2 \propto x^{-\lambda_{eff}}$,
so that $\lambda_{eff}=d\ln F_2/d\ln \frac{1}{x}$.
The value of $\lambda_{eff}$ as an observable at small $x$ has been
discussed by  Navelet et al.\cite{nav1,nav2}. 
The E665 and ZEUS $F_2$ data at fixed $Q^2$ and $x<0.01$ 
have been fitted to the form $A x^{-\lambda_{eff}}$\cite{zeusphenom}.
  \begin{figure}[t]
 \vspace{10.0cm} 
\includegraphics{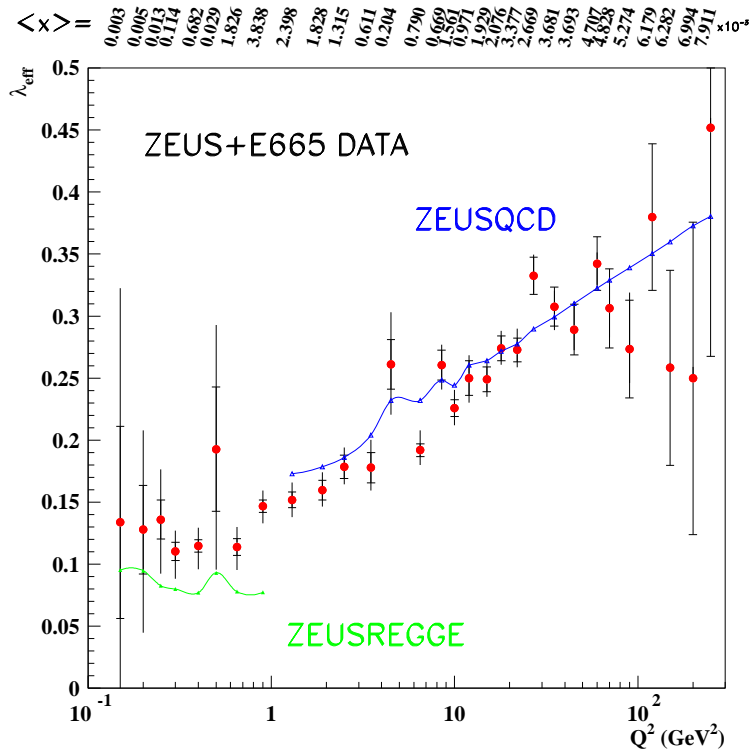}
 \caption{\it
The derivative
$d\ln F_2/d\ln \frac{1}{x}=\lambda_{eff}$ as a function of $Q^2$ calculated
by fitting $F_2=A x^{-\lambda_{eff}}$ to the 
E665 and ZEUS data with $x<0.01$. 
The ZEUSREGGE and ZEUSQCD calculations
are from the Regge and NLO QCD fits{\rm \protect\cite{zeusphenom}}.
The average $x$ values of the various points 
are reported on the top of the Figure.
}
\end{figure}
Fig. 4 shows the measured values of $\lambda_{eff}$ 
as a function of $Q^2$.
Regge phenomenology predicts 
$\lambda_{eff} = \alpha_P (0) - 1 \approx 0.1$ 
independent of $Q^2$, where $\alpha_P (0)$ is the 
intercept of the pomeron trajectory.
Data for $Q^2<1$ GeV$^2$ are consistent with this prediction.
For $Q^2>1$ GeV$^2$, $\lambda_{eff}$ is observed to rise,
reaching approximately  $0.3$ at $Q^2 = 50$ GeV$^2$. 
The rise of $\lambda_{eff}$ with $Q^2$ is described by pQCD\cite{nav2},
in particular by the ZEUS pQCD NLO fit, based on DGLAP evolution equations,
shown in Fig.~4. 
As was the case for the \Ft~measurements, there is no need
in the evolution equations for   
$(\ln \frac{1}{x})^n$ terms in order to  
to describe the logarithmic slope in $x$ of $F_2$.

Even more interesting is the study of $dF_2/d\ln Q^2$. 
At small $x$, 
this derivative is dominated 
by the convolution of the splitting function 
$P_{qg}$ and the gluon density, 
$dF_2/d\ln Q^2\sim \alpha_S P_{qg}\otimes xg$.
As a consequence, $xg$ can be directly related to the measured values of 
$dF_2/d\ln Q^2$\cite{prytz}.
The logarithmic slope $d F_2/d\ln Q^2$ has been calculated using the ZEUS
data by fitting $F_2 = a+b\ln{Q^2}$ in bins of fixed $x$. 
The results\cite{zeusphenom} are shown in Fig. 5.
  \begin{figure}[t]
 \vspace{10.0cm} 
\includegraphics{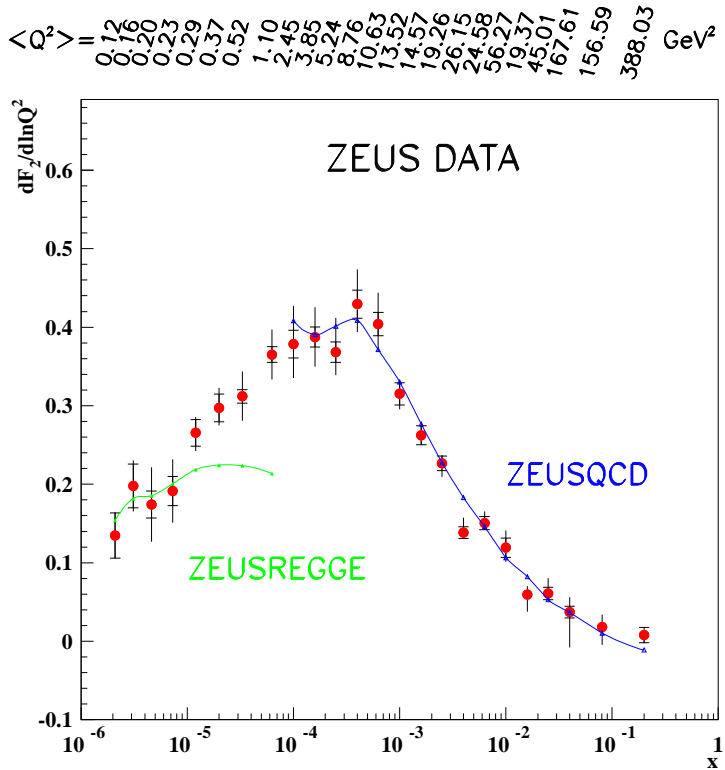}
 \caption{\it
The derivative $d F_2/d\ln Q^2$ as a function of $x$ calculated by fitting
ZEUS $F_2$ data in bins of $x$ to the form $a+b\ln Q^2$.
The ZEUSREGGE and ZEUSQCD calculations
are from the Regge and NLO QCD fits{\rm \protect\cite{zeusphenom}}.
The average $Q^2$ of the various points is shown on the top of the Figure.
}
\end{figure}
For values of $x$ down to $3\times 10^{-4}$, the slopes increase
as $x$ decreases, while at smaller values 
of $x$ and $Q^2$ the slopes decrease.
It should be noted that each point in Fig. 5 corresponds to a different
average value of $Q^2$, as indicated at the top of Fig. 5.
As predicted by pQCD, the `turn over' is not seen
if $dF_2/d\ln Q^2$ is plotted at a fixed value of $Q^2$,
neither in the fixed target data at $Q^2>0.5$ GeV$^2$\cite{caldwell}
nor in the HERA data at $Q^2>3$ GeV$^2$\cite{max}.
Although the `turn over' is partly a kinematic effect 
due to averaging over a 
$Q^2$ range which is different for different $x$ values, it
reflects a smaller rise of the derivatives\cite{golec} 
(i.e. of the gluon density)
for $x\rightarrow 0$ when $Q^2$ decreases below few GeV$^2$.
We will return to this discussion in the next Section.

\subsubsection{\it The pQCD NLO fits: the gluon and 
quark singlet momentum densities}

In order to extract the parton momentum distributions in the proton,
both the H1 and ZEUS collaborations performed a pQCD fit
to the $F_2$ data, solving the DGLAP evolution equations\cite{dglap}
at NLO in the $\overline{MS}$ renormalisation scheme. 
  \begin{figure}[t]
 \vspace{10.0cm} 
\includegraphics{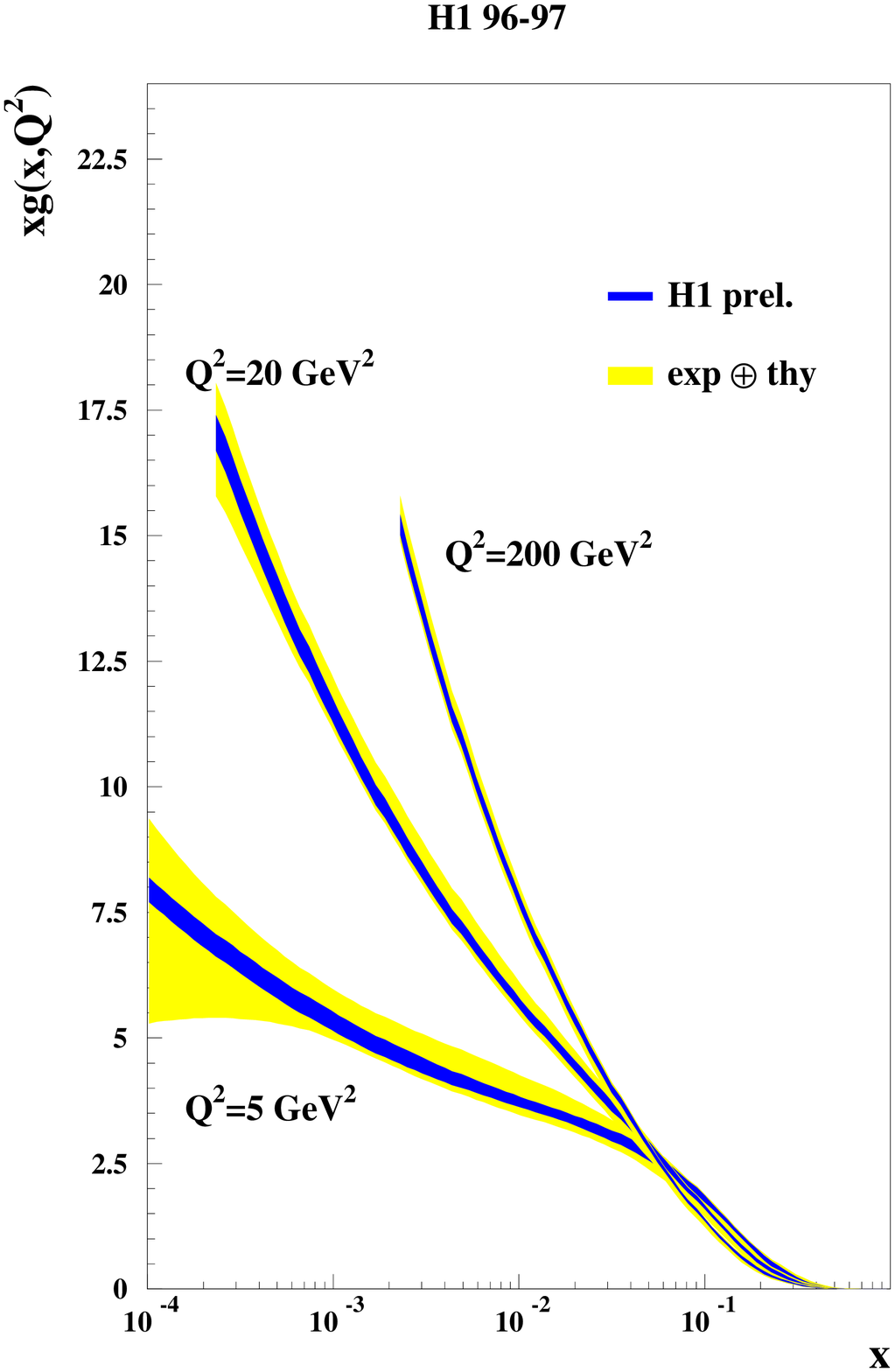}
\includegraphics{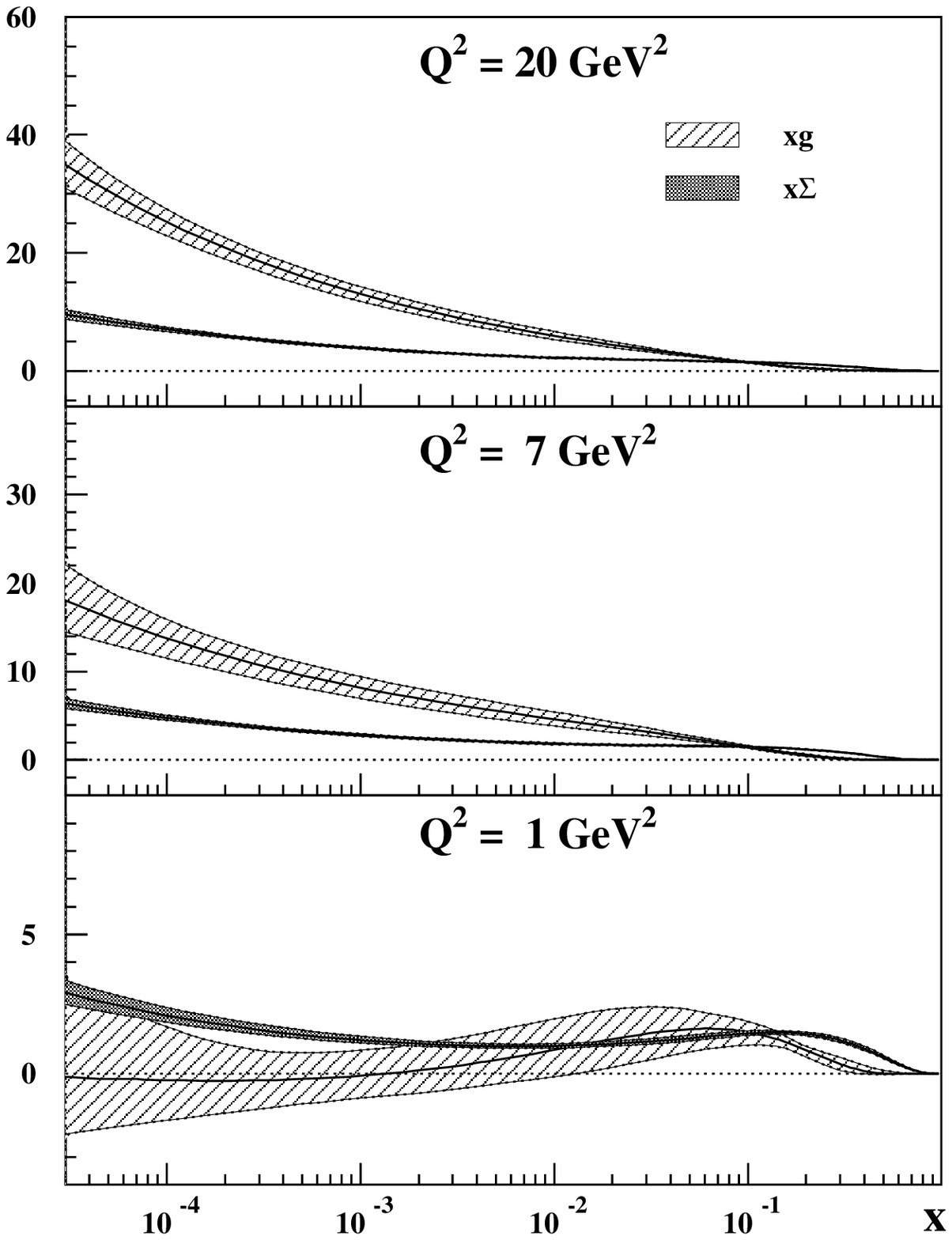}
 \caption{\it
The momentum distributions of partons as obtained 
by the H1 and ZEUS collaborations.
Right: the gluon momentum distribution $xg$ as a function of $x$ at fixed
values of $Q^2 = 5, 20$~and $200$ {\rm GeV}$^2$ from the H1 QCD fit.
Left: the quark singlet momentum distribution, $x\Sigma$ (shaded), and 
the gluon momentum distribution, $xg$ (hatched), as functions of $x$ 
at fixed values of $Q^2 = 1, 7$~and $20$ {\rm GeV}$^2$ 
from the ZEUS QCD fit. 
}
\end{figure}
In both fits, $\alpha_s (M_{Z^0})=0.118$,
the momentum sum rule was applied and 
three light flavours were considered, 
the $c$ and $b$ quarks being generated dynamically 
through boson-gluon fusion (BGF).
The H1 fit\cite{max} used H1 and NMC data at $3.5<Q^2<3000$ GeV$^2$,
while the ZEUS fit\cite{zeusphenom}
used ZEUS, NMC and BCDMS data at $1<Q^2<5000$ GeV$^2$.
The results for the gluon momentum distribution 
$xg(x,Q^2)$ vs. $x$ are shown in Fig. 6
at fixed values of $Q^2$.
Both collaborations measure a strong rise of $xg$ 
for $x\rightarrow 0$ for $Q^2 \gsim 5$ GeV$^2$, 
with the rise increasing with increasing $Q^2$.
In Fig. 6 right, the $xg$ 
obtained by ZEUS is compared to the
quark singlet momentum distribution
$x\Sigma(x,Q^2)=\sum_{f=u,d,s}[xq_f(x,Q^2)+x\bar{q}_f(x,Q^2)]$
obtained with the same fit.
At $Q^2=1$ GeV$^2$,
the sea is still rising at the lowest $x$,
while the gluon, within large uncertainties, is rising much less
and is compatible with zero.
These results are not compatible with the assumption that 
the rise of $F_2$ at $Q^2 \simeq 1$ GeV$^2$
is entirely driven by the increase of the gluon density at small $x$ 
due to parton splitting.

\subsection{Determination of the longitudinal structure function $F_L$
at small $x$}

The longitudinal structure function $F_L$ is predicted by pQCD
to be a function of $F_2$ and $xg$ (see Eq.~\ref{fl}) 
and 
it is expected to give a sizable contribution to the cross section
at large values of $y$.
Therefore a measurement of $F_L$ constitutes 
an important constraint to the theory.
In the measurements of $F_2$ described above, 
$F_L$ was assumed to be equal to the QCD prediction.
A direct measurement\cite{lothar} of $F_L$ requires either running the
HERA collider at different centre-of-mass energies
or using events with initial state QED radiation.
The H1 collaboration extracted $F_L$ from the NC cross section
measurements using the ``subtraction'' method\cite{fl}.
For $Q^2 \ll M_{Z^0}^2$ and  neglecting radiative corrections, 
the NC cross section~(\ref{ncxsect}) can be written:
\[
\frac{d^2\sigma^{ep}_{NC}}{dx\;d\qsd} = \frac{2\pi \alpha^2}{xQ^4}
Y_+ \cdot \sigma_r
\]
where the reduced cross section 
$\sigma_r = F_2-(y^2/Y_+) \cdot F_L$.
Therefore, at large $y$,  $\sigma_r \simeq F_2 - F_L$ and  
$F_L$ may be approximated by $F_2 - \sigma_r$.
The method used by H1 consists in the subtraction
of $\sigma_r$ from $F_2^{QCD}$,
i.e. the result of the  NLO pQCD fit 
to the 1996-97 $F_2$ data at $y<0.35$, for which $F_L \simeq 0$.
The results\cite{max} obtained with this method 
  \begin{figure}[ht]
 \vspace{11.0cm} 
\includegraphics{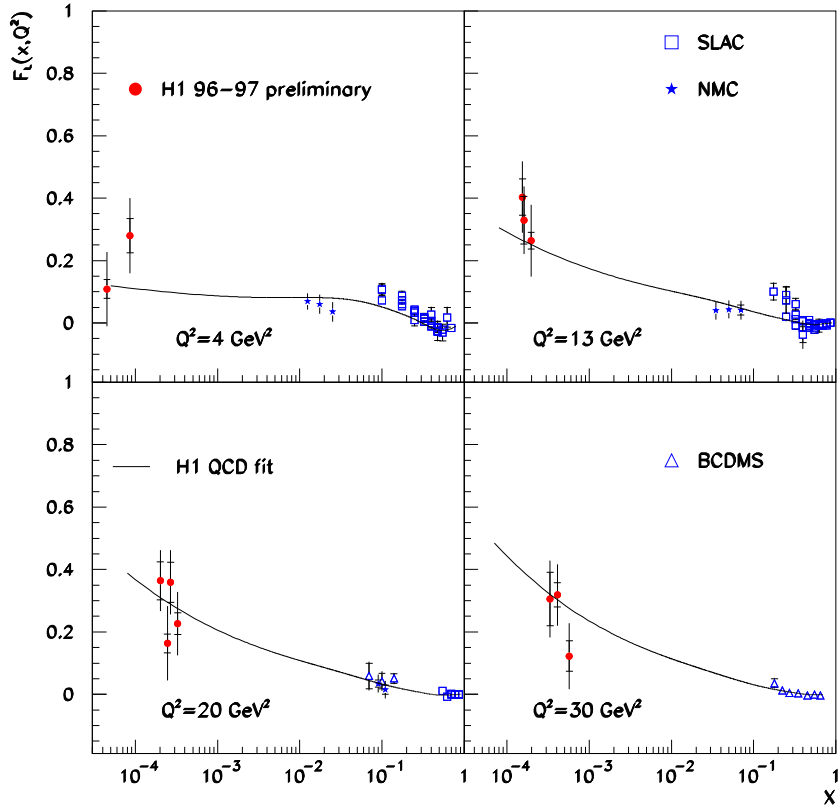}
 \caption{\it
The longitudinal structure function $F_L$ vs. $x$ at fixed $Q^2$ values.
H1, SLAC and NMC data are displayed. The lines represent 
the H1 NLO pQCD fit results.
}
\end{figure}
for $Q^2 > 10$ GeV$^2$
are shown in Fig. 7. 
They rely on the extrapolation of $F_2^{QCD}$ beyond $y=0.35$.
In the same Figure the results obtained
with another method, which relies on assumptions on the
behaviour of the derivative $dF_2/d\ln y$ for $Q^2 < 10$ GeV$^2$,
are also shown.
The extracted $F_L$ is consistent with the QCD prediction.

\subsection{Measurement of the charm structure function $F_2^{c \bar c}$}

Charm production at HERA is expected to be dominated 
by BGF\cite{harris},
i.e. by the gluon density. 
Therefore, given the large $xg$ measured at HERA for $x\rightarrow 0$
in most of the $Q^2$ range,
we expect the charm contribution to $F_2$ to be large.
In analogy with Eq.~(\ref{ncxsect}),
for not too large $Q^2$ and $y$ and neglecting radiative corrections, 
the charm structure function $F_2^{c \bar c}$ is defined as:
\[ 
\frac{d^2\sigma^{c \bar c}}{dx\;d\qsd} = \frac{2\pi \alpha^2}{xQ^4}
Y_+ F_2^{c \bar c} (x,\qsd ).
\]
Experimentally, the cross section for the production of a
$c \bar c$ pair, $\sigma^{c \bar c}$, is extracted from 
the visible cross section for the production of $D^*$ mesons, 
$\sigma^{D^*}$, after correction for the $c \rightarrow D^*$
fragmentation and extrapolation to the full $(\eta, p_T)$ range. 
This measurement is a very effective test of QCD, 
since $F_2^{c \bar c}$ is also calculable
from pQCD knowing $xg$ from the $F_2$ scaling violations
and applying the BGF  NLO calculations\cite{harris}. 
The result of such a calculation 
can be compared to the direct measurement of $F_2^{c \bar c}$.
%
  \begin{figure}[t]
 \vspace{10.0cm} 
\includegraphics{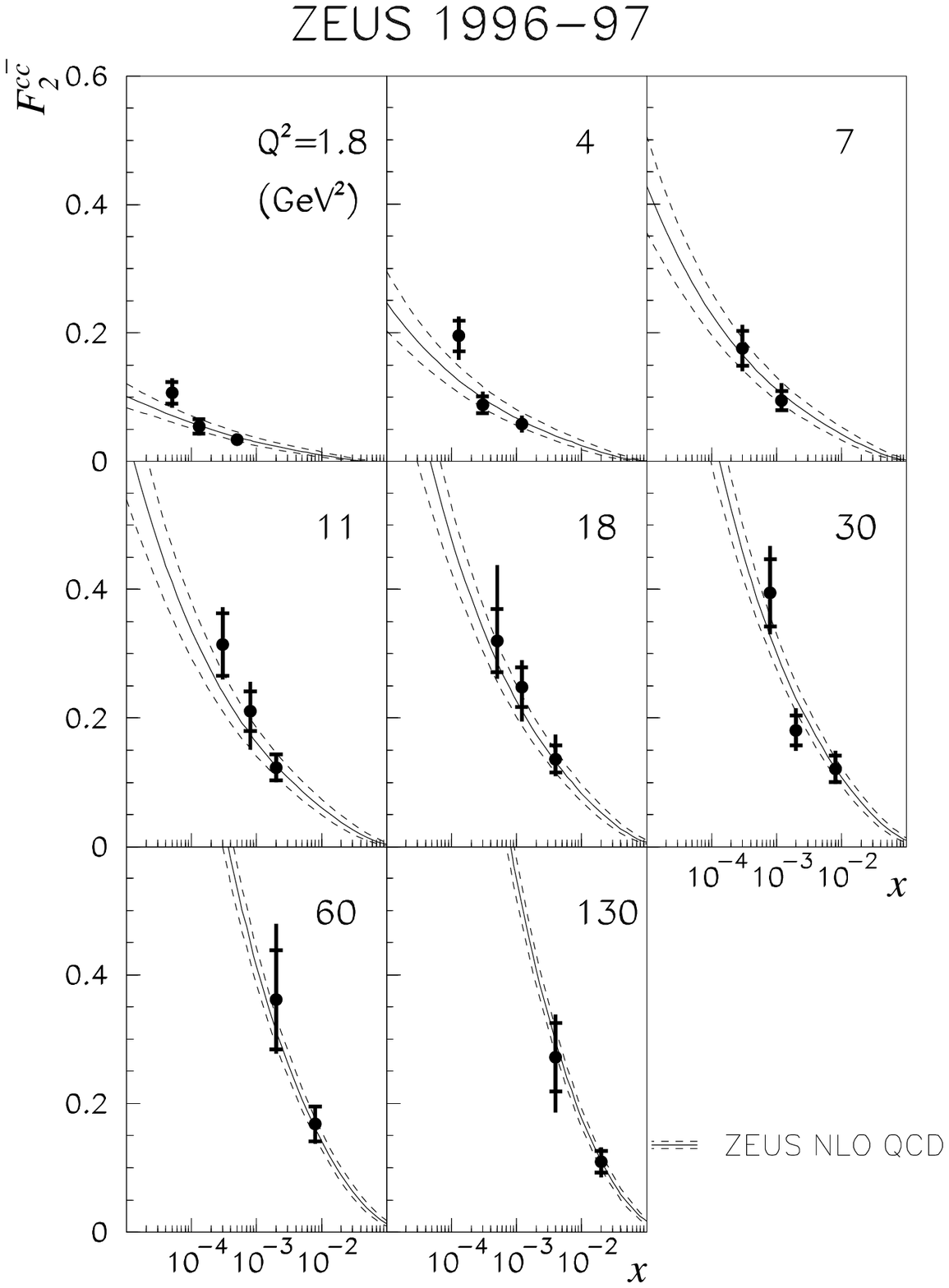}
\includegraphics{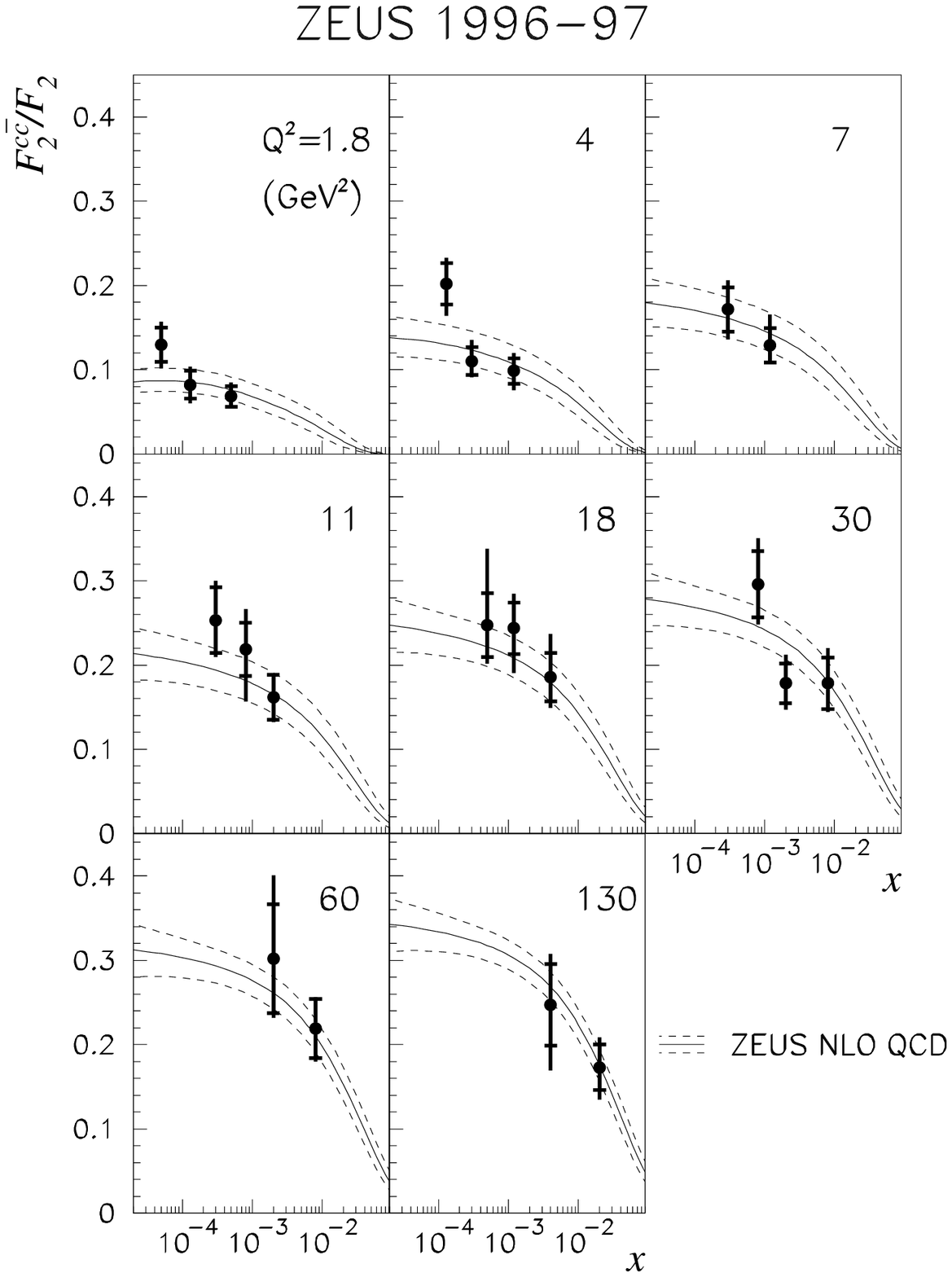}
 \caption{\it
Measurement of $F_2^{c \bar c}$ (left) and $F_2^{c \bar c} / F_2$ (right) 
vs. x in bins of $Q^2$. 
The full lines are the result of the ZEUS pQCD NLO fit, 
while the dashed lines represent the uncertainty of the fit, 
dominated by the variation  by $0.2$ MeV of the charm mass
around the central value $m_c = 1.4$ GeV.
}
\end{figure}
Fig. 8 left, shows the ZEUS measurement\cite{f2charm}
of $F_2^{c \bar c}$ vs. $x$ 
in bins of $Q^2$.
A steep rise of $F_2^{c \bar c}$ is measured for $x \rightarrow 0$,
the rise being more pronounced at large $Q^2$.
The NLO QCD fit is in agreement with the data,
which proves
that the BGF diagram is the dominant mechanism 
for charm production at HERA.
The ratio $F_2^{c \bar c} / F_2$ is shown in Fig. 8 right. 
For $x \rightarrow 0$,
$F_2^{c \bar c}$ rises more rapidly than $F_2$, 
is $\sim 25$ \% of $F_2$ at low $x$ and high $Q^2$
and decreases to 10 \% at $Q^2 = 1.8$ GeV$^2$.
Since $F_2^{c \bar c}$ is dominated by the gluon contribution 
while $F_2$ contains also the sea quarks,
the $F_2^{c \bar c} / F_2$ behaviour\footnote{The effect of the 
charm-mass threshold is negligible at 
the small $x$ values discussed here\cite{jack}} 
is consistent with the hypothesis
that the ratio {\it gluons/(gluons + sea)} 
decreases for $Q^2 \rightarrow 0$,
as has been shown in Sections 3.2.2 and 3.2.3.

\section{{\boldmath $e^\pm p$} cross sections at large {\boldmath $Q^2$}}

\subsection{Neutral currents}

For $Q^2$ beyond a few thousand GeV$^2$, 
the parity violating structure function $xF_3$ becomes sizable
and can no longer be neglected.
  \begin{figure}[t] 
\vspace{9.5cm} 
\includegraphics{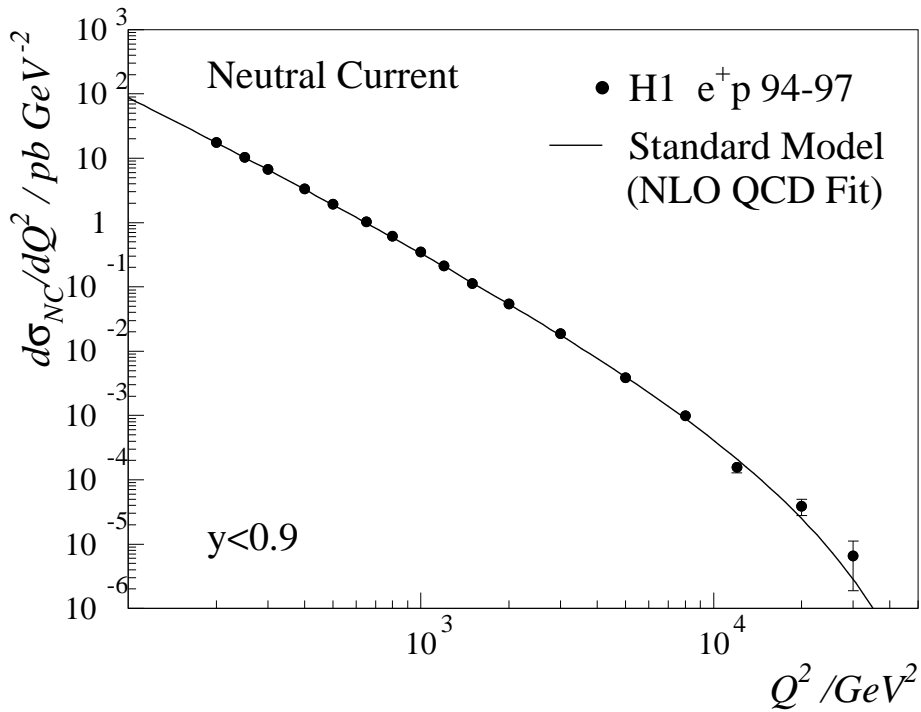}
 \caption{\it
The $d\sigma_{NC}/dQ^2$ for $e^+ p$ scattering at HERA.
The line represents the Standard Model prediction described in the text.
} 
\end{figure}
In this case we will write the NC cross section:
%
\begin{equation}
\frac{d^2\sigma^{e^{\pm} p}_{NC}}{dx\;d\qsd} = \frac{2\pi \alpha^2}{xQ^4}
\left[ Y_+ \tilde{F}_2 (x,\qsd ) \mp Y_{-}x \tilde{F}_3 (x,\qsd )\right]
\label{nc}
\end{equation}
having neglected radiative corrections and  $F_L$.
The structure functions themselves contain 
contributions from virtual photon and $Z^0$ exchange:
\[
\tilde{F}_2  = \Fem + \frac{\qsd}{(\qsd + M_{Z^0}^2)}\Fint
+ \frac{Q^4}{(\qsd + M_{Z^0}^2)^2}\Fwk 
\]
\[
\tilde{F}_3  = \frac{\qsd}{(\qsd + M_{Z^0}^2)} F_3^{int}
+ \frac{Q^4}{(\qsd + M_{Z^0}^2)^2} F_3^{wk}
\]
where the superscripts $em$, $wk$ and $int$ indicate the contributions due 
to photon exchange, $Z^0$ exchange
and $\gamma Z^0$ interference.
The measured\cite{nc} $e^+p$ NC cross section vs.~$Q^2$ is shown in Fig. 9.
The fall of $d\sigma^{e^+ p}_{NC}/dQ^2$ over seven orders of magnitude
constitutes a great success for the Standard Model.
The extrapolation to large $Q^2$ values of 
the NLO pQCD fit, including the EW propagator terms, 
obtained with the data at $Q^2 < 120$ GeV$^2$ 
describes the data well,
proving the validity of the 
theory
in such a wide $Q^2$ range.
Larger luminosities are needed to constrain 
the PDFs at the largest $(x,Q^2)$ values.

The structure function $xF_3$ enters in Eq. (\ref{nc})
with a $-$ or $+$ sign depending if the lepton beam consists of
positrons or electrons, respectively.
Therefore the collection of both $e^-p$ and $e^+p$ data samples
permit a measurement of $xF_3$.
  \begin{figure}[t]
 \vspace{12.0cm} 
\includegraphics{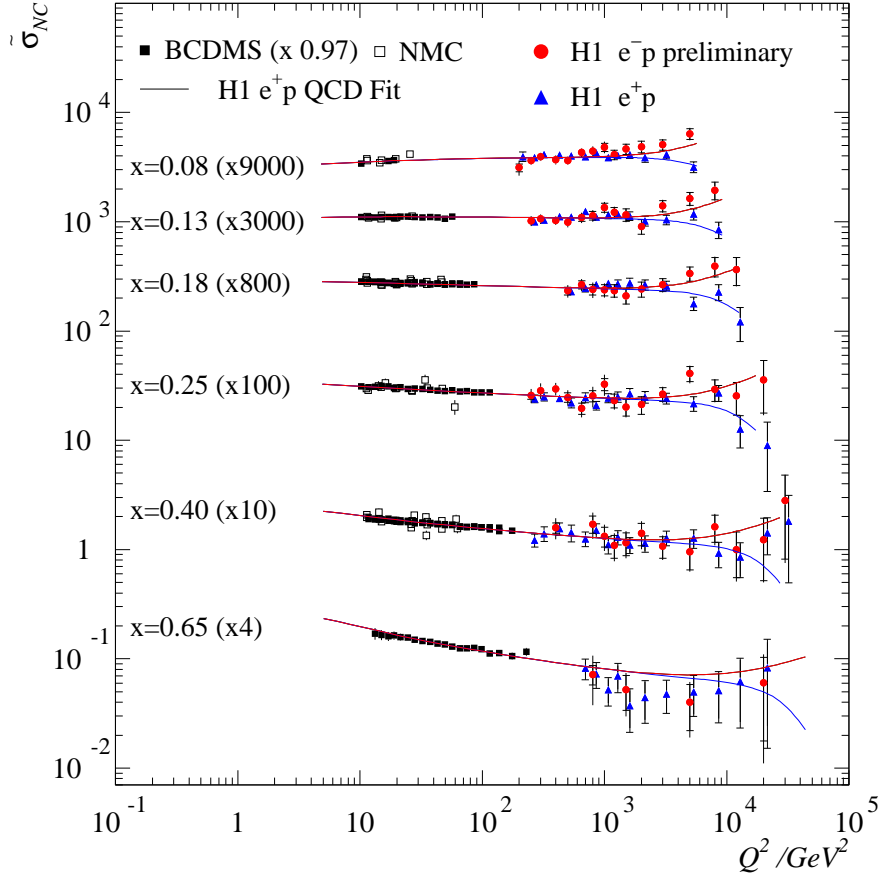}
 \caption{\it
The reduced cross section $\tilde{\sigma}_{NC}$
for $e^+ p$ and $e^-p$ collisions. The effect of the $\gamma Z^0$
interference is evident at the largest $Q^2$ and is in agreement with
the theory.
} 
\end{figure}
The reduced cross section: 
\[ \tilde{\sigma}_{NC} = Y_+ \tilde{F}_2 (x,\qsd ) 
\mp Y_{-}x \tilde{F} _3 (x,\qsd )
\] 
is shown in Fig. 10 for both $e^+ p$ and $e^- p$ data\cite{interference}.
For $Q^2> 3000$ GeV$^2$, the $e^- p$ cross sections 
are larger than the $e^+ p$ ones, 
as expected from the Standard Model $\gamma Z^0$ interference.

\subsection{Charged currents}

The charged current (CC) double-differential cross section can be written:
\begin{equation}
\frac{d^2\sigma^{e^\pm p}_{CC}}{dxdQ^2} = \frac{G_F^2}{2 \pi x} \cdot
\left[ \frac{M_W^2}{Q^2 + M_W^2} \right]^2 \cdot \Phi^\pm_{CC}(x,Q^2)
\label{cc}
\end{equation}
where $G_F$ is the Fermi constant and $M_W$ is the $W^\pm$ boson mass.
In the naive quark-parton model:
\[
\Phi^+_{CC} (x,Q^2) = \bar u + (1-y)^2 \cdot (d+s) 
\]
\[
\Phi^-_{CC} (x,Q^2) = u + (1-y)^2 \cdot (\bar d+ \bar s).
\]
It should be noticed that in the case of $e^- p$ scattering   
the CC cross section is directly sensitive to quarks 
(whereas $\sigma^{e^+p}_{CC}$ is sensitive to antiquarks)
and that in the case of $e^+p$ scattering 
the helicity factor $(1-y)^2$ multiplies the quark densities.
  \begin{figure}[t]
 \vspace{10.0cm} 
\includegraphics{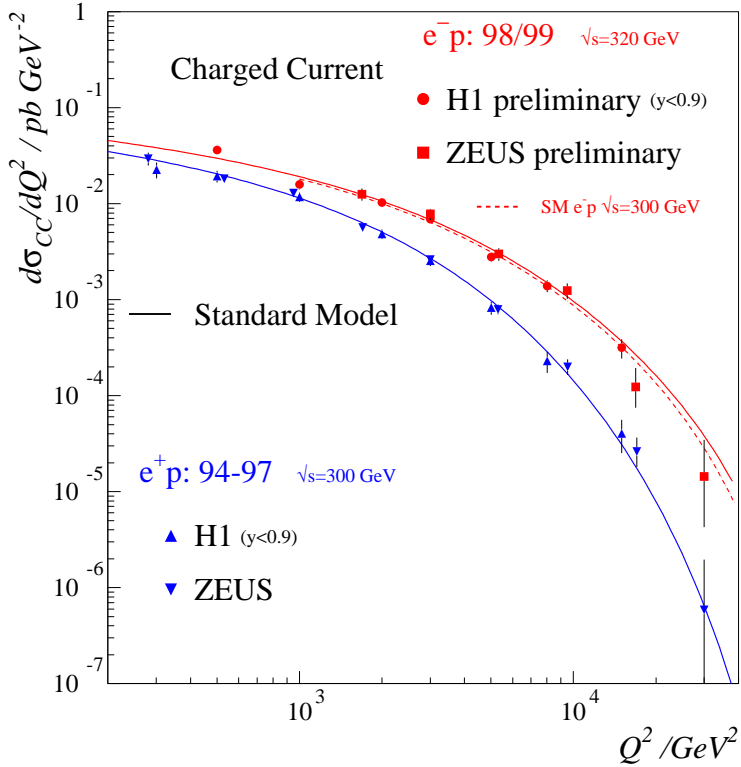}
 \caption{\it
HERA CC differential cross sections vs. $Q^2$ 
for $e^+p$ and $e^-p$ scattering.
The HERA 1994-97 data at $\sqrt{s}=300$ GeV and the 1998-99 data 
at $\sqrt{s}=318$ GeV are plotted.
The lines represent the Standard Model predictions.
}
\end{figure}
For both these reasons
we expect $\sigma^{e^- p}_{CC} > \sigma^{e^+ p}_{CC}$ 
at large $(x,Q^2)$, where valence quarks must dominate.
Fig. 10 shows the HERA CC 
differential cross sections\cite{interference,cch1,cczeus,cceminus} 
as a function of $Q^2$,
both for $e^+p$ and $e^-p$ scattering. 
The difference between 
$\sigma^{e^- p}_{CC}$ and $\sigma^{e^+ p}_{CC}$ increases with $Q^2$,
reaching approximately one order of magnitude at $Q^2 \simeq 10^4$ GeV$^2$.
The Standard Model describes the data well.

\subsubsection{\it Measurements of the W-propagator mass}

Charged current reactions are mediated by the exchange of a virtual $W^\pm$.
It is important to measure the $W$-propagator mass,
i.e. the mass of a {\it spacelike} $W$,
since a deviation from the {\it timelike}-$W$ mass 
measured in $e^+e^-$ and $pp$ collisions
may reveal an anomalous spacelike EW sector.
In Eq.~(\ref{cc}) the absolute magnitude of the cross section 
is given by $G_F$ and the functions $\Phi^\pm$, 
while the cross-section shape
is entirely contained in the propagator term.
  \begin{figure}[t]
 \vspace{12.5cm} 
\includegraphics{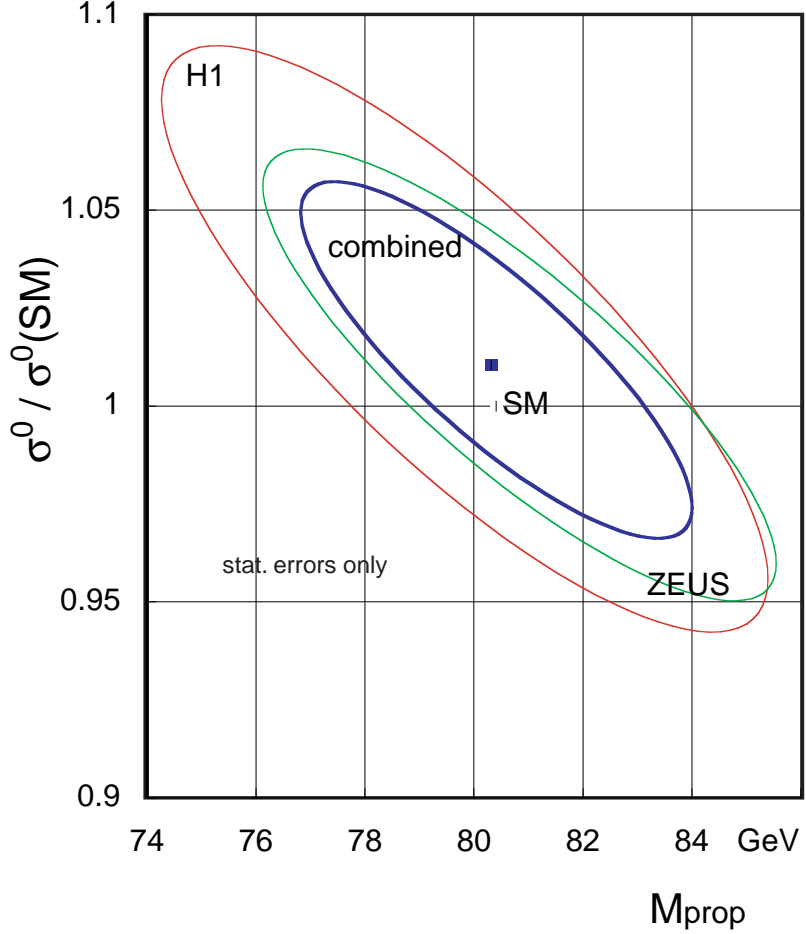}
 \caption{\it
The HERA measurements of the 
CC cross section with respect to the Standard Model,
$\sigma^0 / \sigma^0 (SM) \propto G_{F}$,
and of the $W$-propagator mass. 
The one sigma contours are obtained using only 
the statistical uncertainties.
}
\end{figure}
The H1 and ZEUS collaborations fitted the CC cross section
with two free parameters, the coupling $G_{F}$ and the 
propagator mass $M_{W}$. The results are shown in Fig. 11, 
where the H1 and ZEUS one sigma contour distributions,
and the combined one, are given.
Both collaborations find\cite{cch1,cczeus} the $W$-propagator mass 
in agreement with the timelike-$W$ mass
within a statistical uncertainty of approximately $5$ GeV.
This result proves the universality of the CC interactions
over a wide range of $Q^2$.
The fit has been repeated 
constraining the cross-section normalisation using the precise value
$G_F = (1.16639 \pm 0.00001)\cdot 10^{-5}$ GeV$^2$, measured 
in muon decay. In this case, the results are in agreement with the   
timelike-$W$ mass within $\sim 3 $ GeV $(stat.)$ and 
$\sim 5$ GeV $(stat. \oplus syst.)$.
Finally, 
to exploit the strong dependence of the cross-section 
normalisation ($G_{F}$)
on the shape ($M_{W}$) in a model-dependent fit,
ZEUS used the Standard Model relation:
\[
G_F = \frac{\pi \alpha}{\sqrt{2}} \cdot
\frac{M^2_{Z^0}}{M^2_{W}(M^2_{Z^0}-M^2_{W})} \cdot
\frac{1}{1-\Delta r(M_W)}
\]
where $\Delta r(M_W)$ contains the radiative corrections
to the lowest-order expression for $G_F$ and is a function of
$\alpha$ and the masses of the fundamental bosons and fermions.
Using this relation, the $3 \%$ cross-section  uncertainty is
cast in the uncertainty of a single EW parameter, and the 
uncertainty on the $W$-propagator mass is expected to reduce
by a large factor.
The result of this model-dependent fit is\cite{cczeus}: 
\[
  M_W = 80.50^{+0.24}_{-0.25} (stat.) ^{+0.13}_{-0.16} (syst.)
                                             {\pm 0.31}      (PDF)
                                            ^{+0.03}_{-0.06} ( \Delta M_t, \Delta M_H, \Delta M_Z ) {\mbox{\rm GeV}}. 
  \label{e:SMMW}
\]
where the last uncertainty is obtained by varying the masses of
the top quark, the Higgs and the $Z^0$ bosons.

\subsection{Comparison of NC and CC cross sections}

The Standard Model predicts that, with increasing $Q^2$, NC and CC
should become of equal magnitude. 
This prediction can be verified at HERA,
given the large range in $Q^2$.
Fig. 13 shows the ZEUS and H1 measurements
of the differential $e^-p$ NC and CC cross sections
as a function of $Q^2$.
At $Q^2 \simeq 10^4$ GeV$^2$ the $e^-p$ NC and CC 
cross sections reach similar values.
The Standard Model predictions are in good agreement with the measurements.
We conclude that the unification of charged currents and neutral currents
has been verified at HERA.
  \begin{figure}[ht]
 \vspace{11.0cm} 
\includegraphics{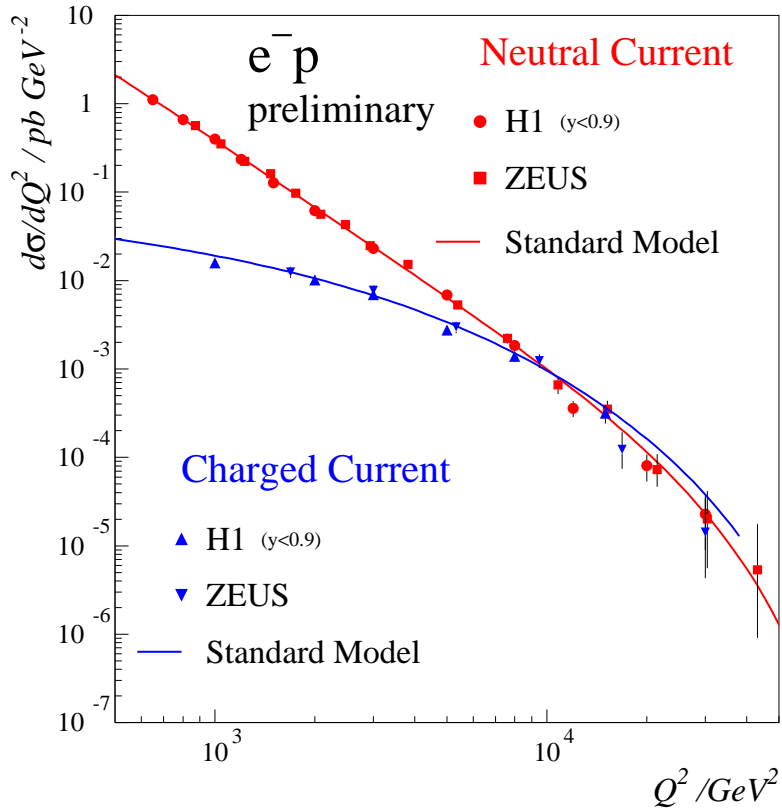}
 \caption{\it
The HERA measurements of NC and CC $e^-p$ cross sections, 
showing their unification at large $Q^2$, 
as predicted by the Standard Model.
}
\end{figure}

\section{Summary and outlook}

The study of structure functions at HERA has reached a mature stage.
The structure function $F_2$ has been measured in a very wide kinematic
range, $10^{-1} < Q^2 < 10^5$ GeV$^2$ and $x$ down to $10^{-6}$.
Precisions of a few percent have been reached 
in a large fraction of the above range.
At the lowest $Q^2$ measured, $F_2$ can be described 
by Regge phenomenology.
For $Q^2$ above few GeV$^2$, 
the region where pQCD is expected to be applicable, 
the DGLAP evolution works well 
and no need is found for the BFKL $(\ln \frac{1}{x})^n$ correction terms.
In this region $F_2$ rises strongly towards small $x$; 
hints have been found that at small $Q^2$ this rise 
may be driven by sea quarks, and not by gluons.
The charm structure function $F_2^{c \bar c}$ is measured 
to be a substantial contribution to $F_2$, 
reaching 25\% at small $x$ and large $Q^2$.
Indirect determinations of the longitudinal structure function $F_L$
agree with the QCD expectation.

The first measurements of the NC and CC cross sections 
at very large $Q^2$ have been made in $e^\pm p$ scattering. 
The Standard Model predictions agree with the data.
In NC interactions,  $\gamma Z^0$ interference
has been observed, while in CC interactions 
the $W$-propagator mass has been measured to be consistent
with the timelike-$W$ mass within a few GeV.
The unification of NC and CC interactions at very large $Q^2$
has been measured in $e^- p$ scattering.

Overall, the Standard Model is found to be in good agreement
with the measurements based on approximately $100~pb^{-1}$ of data 
collected at HERA by each collider experiment.
The plans for the future are 
to increase the integrated luminosity by an order of magnitude.
In Summer 2000 HERA will be shut down for nine months.
During this period, superconducting magnets will be installed
inside the H1 and ZEUS detectors to achieve stronger beam focusing
at the interaction points and 
obtain an increase of a factor of five 
in specific luminosity.
The plans are to run for at least five more years to integrate 
$\sim 1~fb^{-1}$ of data,
which will give full access to the electroweak physics programme 
and to the search for physics beyond the Standard Model
at HERA.

\section{Aknowledgements}
I wish to thank the organisers of the workshop for providing 
once again the optimal conditions to enjoy life and physics
at the same time. 
I am indebted to the many ZEUS and H1 colleagues
from which I have learned much.
I particularly thank Brian Foster for carefully 
reading and commenting on the manuscript
and Jack Smith for elucidating theoretical aspects 
of charm production at HERA.


\begin{thebibliography}{99}
\setlength{\itemsep}{-2pt}

\bibitem{uhn}
R.~Gandhi {\it et al.},
Phys. Rev. {\bf D 58}, 093009 (1998); \\
J.~Kwiecinski {\it et al.},
hep-ph/0004109.
\bibitem{mandy} A.M.~Cooper-Sarkar {\it et al.},
Int. J. Mod. Phys. {\bf A13}, 3385 (1988), and ref. therein.
\bibitem{sat} L.V.~Gribov  {\it et al.},
Phys. Rep. {\bf 100}, 1 (1983).
\bibitem{bfkl} 
V.S.~Fadin  {\it et al.},
Phys. Lett. {\bf B422}, 287 (1998); \\
G.~Altarelli {\it et al.},
hep-ph/0001157 and ref. therein.
\bibitem{ew} R. Cashmore  {\it et al.}, 
Proc. of the DESY workshop on Future Physics at HERA, 
ed. G.~Ingelman {\it et al.}, 127 (1996),
http://www.desy.de/~heraws96.
\bibitem{bpt}
ZEUS Collab., J.~Breitweg  {\it et al.},
DESY 00-071 (May 2000), submitted to Phys. Lett. B. 
\bibitem{max} Max Klein, Talk given at the XIX International 
Symposium on Lepton and Photon Interactions at High Energies, 
Stanford, August 1999, hep-ex/0001059. 
\bibitem{niels} Niels Tuning, Talk given at the DIS2000
Workshop on DIS, Liverpool, April 2000.
\bibitem{dglap}    
G.~Altarelli and G.~Parisi, Nucl. Phys. {\bf B126}, 298 (1977); \\
V.~Gribov and L.~Lipatov, Sov. Jour. Nucl. Phys. {\bf 15}, 438 (1972); \\
L.~Lipatov, Sov. Jour. Nucl. Phys. {\bf 20}, 94 (1975); \\
Y.~Dokshitzer, Sov. Phys. JETP {\bf 46}, 641 (1977).
\bibitem{bartels}J.~Bartels {\it et al}, 
Proc. of the DESY Workshop Physics at HERA, 
ed. W. Buchm\"uller and G.~Ingelman, 193 (1991).
\bibitem{nav1} H.~Navelet, R.~Peschanski and  S.~Wallon, 
Mod. Phys. Lett {\bf A36}, 3393  (1994).
\bibitem{nav2} H.~Navelet et al., Phys. Lett. {\bf B385} (1996) 357.
\bibitem{zeusphenom} ZEUS Collab., J.~Breitweg {\it et al.}, 
Eur. Phys. J. {\bf C7}, 609 (1999).
\bibitem{prytz} K. Prytz, Phys. Lett. {\bf B311}, 286 (1993); 
ibid {\bf B 332}, 393 (1994).
\bibitem{caldwell} 
A.D.~Martin {\it et al.}, Eur. Phys. J. {\bf C4}, 463 (1998).
\bibitem{golec} K.~Golec-Biernat and M.~W\"unsthoff,
Phys. Rev. {\bf D60}, 114023 (1999) and ref. therein.
\bibitem{lothar} L.A.T.~Bauerdic {\it et al.}, 
Proc. of the DESY workshop on Future Physics at HERA, 
ed. G.~Ingelman {\it et al.}, 77 (1996),
http://www.desy.de/~heraws96.
\bibitem{fl} H1 Collab., C. Adloff {\it et al.},
Phys. Lett. {\bf B393}, 452 (1997). 
\bibitem{harris} B.W.~Harris and J.~Smith,
Phys. Rev. {\bf D57}, 2806 (1998).
\bibitem{f2charm} ZEUS Collab., J.~Breitweg {\it et al.}, 
Eur. Phys. J {\bf C12}, 35 (2000).
\bibitem{jack}Jack Smith,
Proc. of the Workshop New Trends in HERA Physics, 
Ringberg, May 1997, hep-ph/9708212.
\bibitem{nc} H1 Collab., C.~Adloff  {\it et al.},
Eur. Phys. J. {\bf C13}, 609 (2000), \\
ZEUS Collab., J.~Breitweg {\it et al.},
Eur. Phys. J. {\bf C11} 3, 427 (1999). 
\bibitem{interference} H1 Collab., C.~Adloff  {\it et al.},
contribution n. 157b, EPS99, Tampere, Finland, July 1999.
\bibitem{cch1} H1 Collab., C.~Adloff  {\it et al.},
contribution n. 157ai, EPS99, Tampere, Finland, July 1999. 
\bibitem{cczeus}
ZEUS Collab., J.~Breitweg {\it et al.},
Eur. Phys. J. {\bf C12} 3, 411 (2000).
\bibitem{cceminus} ZEUS Collab., J.~Breitweg {\it et al.},
contribution n. 558, EPS99, Tampere, Finland, July 1999.


%
\end{thebibliography}
\end{document}